\documentclass{aa}  
\usepackage{graphicx}
\usepackage{txfonts}
\usepackage{natbib}
\bibpunct{(}{)}{;}{a}{}{,}
\usepackage{amsmath}
\usepackage{array}
\usepackage{float}
\usepackage{import}
\usepackage{bibunits}
\usepackage[version=4]{mhchem}
\usepackage{siunitx}

\begin{document}

   \title{Electron processing of glycine: evidence for peptide-forming condensation reactions}


   \author{A.~T.~Hopkinson
          \inst{1}
          \and
          D.~V.~Mifsud
          \inst{2}
          \and
          P.~Herczku
          \inst{2}
          \and
          Z.~Juhász
          \inst{2}
          \and
          B.~Sulik
          \inst{2}
          \and
          N.~J.~Mason
          \inst{2,3}
          \and
          N.~C.~Jones
          \inst{4}
          \and
          S.~V.~Hoffmann
          \inst{4}
          \and
          A.~Cassidy
          \inst{1}
          \and
          L.~Hornekær
          \inst{1}
          \and
          S.~Ioppolo
          \inst{1}
          }

   \institute{Centre for Interstellar Catalysis, Department of Physics and Astronomy, Aarhus University, DK 8000 Aarhus, Denmark\\
              \email{a.hopkinson@phys.au.dk}
              \email{s.ioppolo@phys.au.dk}
              \and
             HUN-REN Institute for Nuclear Research (Atomki), Debrecen H-4026, Hungary
             \and
             Centre for Astrophysics and Planetary Science (CAPS), School of Engineering, Mathematics and Physics, University of Kent, Canterbury CT2 7NH, United Kingdom
             \and
             ISA, Department of Physics and Astronomy, Aarhus University, DK 8000 Aarhus, Denmark
}
   \date{}

 
  \abstract
   {The discovery of amino acids on asteroids, including samples returned from asteroid Bennu, and the detection of glycine on comet 67P/Churyumov–Gerasimenko (67P) support the hypothesis that glycine can form in extraterrestrial environments. Laboratory experiments further demonstrate that glycine can be produced through both energetic and non-energetic processing of interstellar ice analogues.}
   {In the interstellar medium, glycine is exposed to various forms of energetic processing, including cosmic rays and the secondary electrons they generate. Such processing has been shown to produce complex organic molecules (COMs), including peptides. This work investigates whether peptide-forming chemistry is a general consequence of energetic processing by examining the effects of energetic electron irradiation on glycine.}
   {Films of glycine and its partially deuterated isotopologue, glycine-d$_{3}$, were grown in two ultrahigh-vacuum (UHV) chambers and irradiated with 1-2\,keV electrons. The irradiated samples were subsequently characterised using infrared (IR) spectroscopy, temperature-programmed desorption (TPD), and vacuum-ultraviolet (VUV) spectroscopy.}
   {Peptide formation through condensation reactions would produce water as a secondary product, and IR spectra obtained following irradiation of glycine-d$_{3}$ showed evidence of \ce{D2O} formation. This interpretation was supported by TPD measurements, which revealed \ce{H2O} formation for glycine and \ce{H2O}, \ce{HDO}, and \ce{D2O} for glycine-d$_{3}$. VUV spectra of irradiated glycine showed the formation of decomposition products, including \ce{CO} and \ce{CO2}, as well as \ce{H2O}. Following heating, a non-volatile residue remained on the surface, consistent with the formation of larger organic species.}
    {Electron irradiation of glycine produces reaction products similar to proton irradiation, suggesting that the resulting chemistry is not strongly dependent on the energy source. These findings support the possibility that larger organic molecules, including peptides, can form in a wide range of extraterrestrial environments. Such molecules may subsequently be incorporated into small Solar System bodies and delivered to planetary surfaces, potentially contributing to prebiotic chemistry and the emergence of life.}


   \maketitle
%

 \section{Introduction}

One group of complex organic molecules (COMs) considered essential to the origin of life is amino acids. Amino acids can react to form peptides—chains of amino acids that can subsequently form proteins which are fundamental to life as we know it \citep{doi:10.1021/cr2004844}. Peptides play critical roles in cellular architecture and can also assemble into supramolecular structures. The simplest amino acid is glycine (\ce{NH2CH2COOH}), and although glycine has not yet been detected unambiguously in the interstellar medium (ISM), glycine and other amino acids that are thought to have an extraterrestrial origin have been identified in carbonaceous chondrites \citep{10.1039/c2cs35109a}. More recently, samples returned from asteroid Bennu by the \textit{OSIRIS-REx} mission revealed 33 amino acids, including 14 of the 20 standard proteinogenic amino acids such as glycine \citep{2025NatAs...9..199G}. Glycine was also detected on 67P/Churyumov-Gerasimenko (67P) by ESA’s \textit{Rosetta} mission \citep{2019A&A...630A..32H, 2015Sci...347a0628C}. Cometary material is thought to have undergone relatively little thermal processing since its incorporation into the Solar System, allowing it to preserve a record of primitive material that may include an interstellar chemical heritage \citep{2018SSRv..214...36A, 2020SSRv..216..102R}. Laboratory experiments simulating ISM conditions have explored pathways for glycine formation. Glycine has been shown to form through energetic processing of interstellar ice analogues, including ultraviolet irradiation \citep{Uvamino,anotherUVamino}, electrons \citep{2013ApJ...765..111K,Tushar}, as well as through non-energetic reactions in the icy mantles of dust grains under conditions relevant to dense molecular clouds \citep{2021NatAs...5..197I}. Together, these laboratory results and the detection of glycine in primitive Solar System bodies support a scenario in which glycine can form during the dense-cloud stage and continue to be produced throughout subsequent stages of stellar evolution. It may then be incorporated into planetesimals and other small bodies and ultimately delivered to planetary surfaces, where it could contribute to prebiotic chemistry.

Glycine present in the interstellar medium is therefore expected to be exposed to energetic processing throughout the evolution of dense molecular clouds. Previous studies have investigated the effects of UV photons, electrons, and protons on the stability and survivability of glycine. These experiments have shown that energetic processing leads to glycine destruction and the formation of smaller molecular products, including \ce{CO}, \ce{CO2}, \ce{CN-}, and \ce{OCN-} \citep{mate,2014FaDi..168..267M,2014MNRAS.441.3209P}. In several IR studies, amide absorption bands were also detected in the residues remaining after irradiation \citep{mate,protom,2014MNRAS.441.3209P}. The presence of these bands is consistent with the possible formation of peptide bonds and larger peptide-like species.

More recently, energetic processing of glycine with 1\,MeV and 10\,keV protons was shown to produce larger COMs, including peptides \citep{2026NatAs..10..531H}. This chemistry was proposed to be driven primarily by low-energy secondary electrons generated during proton irradiation \citep{2019CSRev..48.2293A}. If low-energy secondary electrons are indeed responsible for initiating the relevant reaction pathways, similar chemistry should occur irrespective of the primary energetic source, provided that secondary electrons are produced. To test this hypothesis, the present study investigates the irradiation of glycine with 1-2\,keV electrons. Particular attention is given to the formation of water during irradiation, as water production would be consistent with peptide formation through condensation reactions, as previously demonstrated by \cite{2026NatAs..10..531H}. Water provides a useful indirect probe of this chemistry because it can be detected more readily than newly formed peptides and other larger COMs. An example of a condensation reaction expected to occur during electron irradiation of glycine is shown in Eq.~\ref{pepform}.

\begin{equation}\label{pepform}
     \ce{R_1-COOH} + \ce{NH2-R_2} \rightarrow \ce{R_1-CONH-R_2} + \ce{H2O}
\end{equation}
\begin{sloppypar}

Glycine exhibits strong, broad IR absorption in the $3400-2900$\,cm$^{-1}$ region, which overlaps with the \ce{O-H} stretching band of water \citep{OD}. Consequently, changes in this spectral region cannot be unambiguously attributed to water formation. To overcome this limitation, partially deuterated glycine-d$_{3}$ (\ce{ND2CH2COOD}) was also irradiated. In glycine-d$_{3}$, the hydrogen atoms associated with the amino and carboxyl functional groups are replaced by deuterium, while the hydrogen atoms bound to the central alpha-carbon remain unchanged. Because peptide formation through condensation involves the amino and carboxyl functional groups, this reaction is expected to release fully deuterated water, \ce{D2O}, as illustrated in Eq.~\ref{pepformD3}. The \ce{O-D} stretching mode of \ce{D2O} occurs in the $2700-2400$\,cm$^{-1}$ region \citep{OD}. Glycine-d$_{3}$ has fewer and weaker overlapping IR absorption features in this region, allowing \ce{D2O} formation to be identified more readily by IR spectroscopy.

\end{sloppypar}
\begin{equation}\label{pepformD3}
     \ce{R_1-COOD} + \ce{ND2-R_2} \rightarrow \ce{R_1-COND-R_2} + \ce{D2O}
\end{equation}

Electron irradiation of glycine was also performed in a separate experimental chamber, where the processed material was characterised using VUV spectroscopy. Compared with IR spectroscopy, VUV spectroscopy provides a more direct means of identifying water formation because the electronic absorption features of water are more clearly separated from those of glycine. Detection of water following electron irradiation of glycine therefore provides an independent test of whether the same reaction pathways observed for partially deuterated glycine-d$_{3}$ also occur for non-deuterated glycine. If electron irradiation produces water together with peptides and other COMs, this would support the hypothesis that the resulting chemistry is largely independent of the primary source of energetic processing. Energetic processing of amino acids may therefore provide a general pathway towards the formation of larger COMs and peptides across a wide range of interstellar environments.

\section{Methods}

Experiments were conducted using two complementary experimental setups. Infrared (IR) spectroscopy measurements were performed with the Ice Chamber for Astrophysics–Astrochemistry (ICA), an ultrahigh-vacuum (UHV) chamber located at the HUN-REN Institute for Nuclear Research (Atomki) in Debrecen, Hungary. Vacuum-ultraviolet (VUV) spectroscopy measurements were carried out at the Aarhus University ultraviolet (AU-UV) beamline on ASTRID2. A summary of the experimental conditions is provided in Table~\ref{expover}. 

\begin{table}[h]
\caption{A list of experiments performed to investigate the energetic processing of glycine with electrons.}              
\label{expover}  
\centering                                      
\begin{tabular}{c c c c}          
\hline\hline                        
 
UHV setup & Measurement type & e$^-$ energy & Molecule \\  
\hline
    \\[-0.8em]
     ICA & IR and QMS & 2~keV & glycine\\ 
     ICA & IR and QMS & 2~keV & glycine-d$_{3}$\\ 
     AU-UV & VUV & 1~keV & glycine\\ 
\hline
\end{tabular}
\end{table}

ICA consists of a UHV-compatible spherical chamber with an inner diameter of 160\,mm. It has a base pressure of $1\times10^{-8}$\,mbar, which is achieved with the use of a roughing pump and a turbo molecular pump, and when the sample is at cryogenic temperatures, the base pressure drops to $1\times10^{-9}$\,mbar. The centre of the chamber is occupied by the sample holder. This consists of three IR-transparent zinc selenide (\ce{ZnSe}) windows that are mounted into a copper sample holder, which is surrounded by a heat shield. Each window allows for a separate irradiation to occur. The sample can be moved and rotated using a rotational stage and manipulator. The sample holder is connected to the cold finger of a closed-cycle helium cryostat (Leybold Coolpower 7/25 with a Leybold Coolpak 4000 compressor unit), allowing the sample to be cooled down to 20\,K. The temperature was measured and controlled using a cryogenic temperature controller (Lake Shore model 335), and the temperature could be changed by heating the sample with an internal 25\,$\Omega$/100 W cartridge heater (HTR-25-100). Glycine molecules were introduced to the chamber through the use of an effusive evaporator. The chamber is equipped with a Fourier transform infrared (FTIR) Bruker V70v spectrometer with mercury-cadmium-telluride (MCT) detector in transmission mode \citep{2021RScI...92h4501H}. The detector was used with a resolution of 1 cm$^{-1}$, each spectrum consists of 128 coadded scans, with a wavelength range from 3750 to 750 cm$^{-1}$. A quadrupole mass spectrometer (QMS) with a mass range of 1 to 300~amu is also mounted on the chamber to measure species that desorb from the surface. The QMS was used in TPD measurements, while noting the start and end times and temperatures and ramping the temperature up linearly at 2 K/min. The chamber was also equipped with a Kimball ELG-2A electron gun that was used to process the deposited molecules with 2\,keV electrons with a flux of $4.5\times10^{13}$\,\ce{e-}\,cm$^{-2}$ \citep{icaelectornflux1,icaelectornflux2}. 

VUV spectroscopy experiments were performed at the AU-UV end station at the ASTRID2 synchrotron facility located at Aarhus University. The AU-UV beam-line produces photons in the VUV and UV-visible regions, between 110 and 700\,nm \citep{jin}. Through the use of a toroidal grating monochromator, the beamline produces tunable VUV photons with a photon flux of $1\times10^{10}$\,photons\,s$^{-1}$ 100\,mA$^{-1}$. This low flux results in negligible VUV photoprocessing of the deposited molecules, enabling VUV photons to be used as a spectroscopic tool only. For this experiment, a wavelength range of 120 to 325 nm was used for measurement with a step size of 1 nm. The UHV chamber mounted on the AU-UV beamline during these experiments consists of a cubic chamber operating at a base pressure of approximately $1\times10^{-9}$\,mbar, maintained using turbomolecular pumps \citep{UVwaterCOandCO2}. The substrate was a magnesium fluoride (\ce{MgF2}) window, transparent in the VUV wavelength range. The substrate was mounted in a copper sample holder and was in thermal contact with a cold finger, cooled to 20\,K using a closed-cycle helium cryostat. To process molecules deposited on the substrate, a SPECS EQ 22/35 electron gun mounted on the chamber allowed for the bombardment of the sample with 1\,keV electrons with a flux of $2.0\times10^{13}$\,\ce{e-}\,cm$^{-2}$ \citep{UVwaterCOandCO2}.

For both experimental setups, the same general procedure was used to investigate the energetic processing of glycine and is therefore described only once. First, background spectra were recorded at the base temperature of 20\,K to characterise the light-source profile and the absorption properties of the clean substrate. Glycine was then deposited onto the 20 K substrate using an evaporation source heated under vacuum to 125\,$^\circ$C, with the substrate positioned normal to the glycine source. Spectra of the as-deposited glycine were then recorded. The glycine samples were subsequently irradiated with electrons in a series of sequential steps, as summarised in Table~\ref{irrsteps}. Spectra were recorded after each irradiation step to monitor the evolution of the sample. Once the irradiation sequence was complete, a final spectrum was acquired before the sample was heated, either to 125 K and then 200K for the VUV spectra or during a TPD measurement for the IR spectra, to desorb volatile irradiation products, including \ce{CO2} and \ce{H2O}. Removal of these species reduced spectral congestion caused by overlapping absorption bands and enabled improved spectroscopic characterisation of the less volatile material remaining on the surface. In experiments performed in the ICA chamber, species desorbing during heating were additionally monitored using a QMS as the surface was ramped to 300 K. 

\begin{table}[h]
\caption{Cumulative electron fluence applied during the sequential irradiation steps in both experiments.}              
\label{irrsteps} 
\centering                                      
\begin{tabular}{c c}          
\hline\hline                        
ICA \ce{e-} fluence (\ce{e-}\,cm$^{-2}$) & AU-UV \ce{e-} fluence (\ce{e-}\,cm$^{-2}$) \\
\hline
    \\[-0.8em]
     $2.25\times10^{14}$ & $2.00\times10^{14}$ \\
     $4.50\times10^{14}$ & $1.40\times10^{15}$ \\
     $1.35\times10^{15}$ & $1.34\times10^{16}$ \\
     $2.70\times10^{15}$ & $3.60\times10^{16}$ \\
     $1.35\times10^{16}$ &  \\
     $2.70\times10^{16}$ &  \\
     $8.10\times10^{16}$ &  \\
     $1.62\times10^{17}$ &  \\
\hline
\end{tabular}
\end{table}

\section{Results}
\subsection{IR spectroscopic measurements}\label{irsection}

\begin{figure}[h]
     \resizebox{0.9\hsize}{!}{\includegraphics{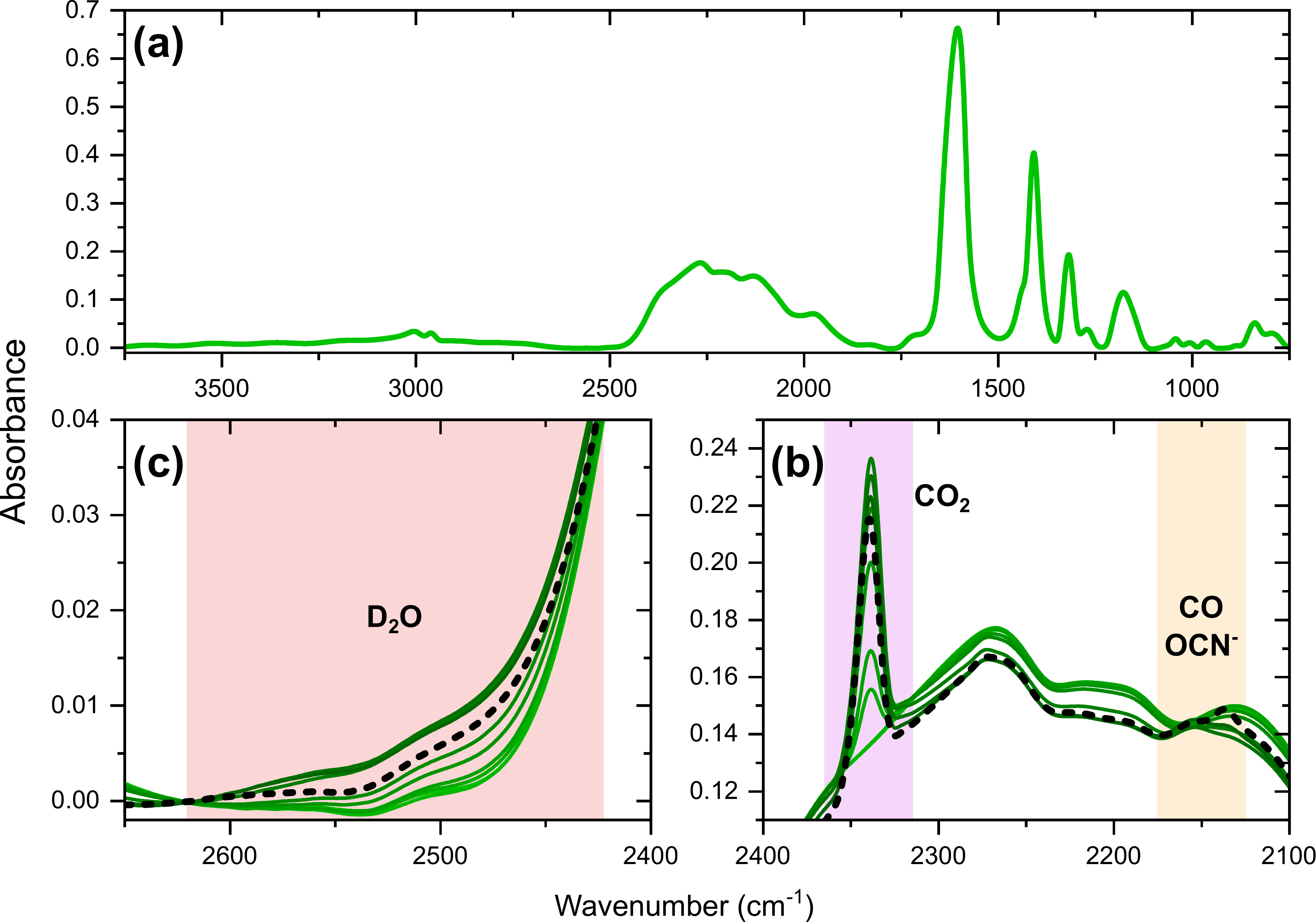}}
     \centering
     \caption{IR spectra of glycine-d$_3$ recorded as a function of 2 keV electron fluence, with darker green traces corresponding to higher fluences up to a maximum of $1.6\times10^{17}$\,\ce{e-}\,cm$^{-2}$ shown in black dashes. (a) FTIR spectrum of partially deuterated glycine-d$_3$ at 20\,K, present predominantly in its zwitterionic form. (b) Spectra over the $2400-2100$\,cm$^{-1}$ region, highlighting absorption features associated with \ce{CO2} (purple highlight), \ce{CO}, and \ce{OCN-} (orange highlight). (c) Spectra over the $2700-2400$\,cm$^{-1}$ region, encompassing the \ce{O-D} stretching band expected for \ce{D2O} (red highlight).}
     \label{eirr}
 \end{figure}

Figure \ref{eirr}(a) shows the IR transmission spectrum of unprocessed, partially deuterated glycine-d$_{3}$ deposited at 20\,K. The principal absorption bands observed in Fig.~\ref{eirr}(a), together with their vibrational assignments and corresponding literature values, are summarised in Table~\ref{partdeutvalues}.

The spectrum in Fig.~\ref{eirr}(a) exhibits a weak shoulder at 1715\,cm$^{-1}$ and a weak absorption band at 1269\,cm$^{-1}$, both characteristic of neutral glycine. These features are assigned to the \ce{C=O} stretching mode and the combined \ce{CO} stretching and \ce{COD} scissoring modes of glycine-d$_{3}$, respectively \citep{glyenviron}. Their relatively low intensities compared with the dominant absorption features indicate that the deposited glycine-d$_{3}$ is present predominantly in its zwitterionic form.

\begin{table*}[t]
\caption{Assignment of the dominant absorption bands observed in the IR spectrum of deposited partially deuterated glycine-d$_3$ shown in Fig.~\ref{eirr}(a).}              
 \label{partdeutvalues}
\centering                                      
\begin{tabular}{c c c}          
\hline\hline                        
Vibrational assignment & Experimental position (cm$^{-1}$) & Literature position (cm$^{-1}$)$^a$ \\ 
      &  & \\
\hline
    \\[-0.8em]
     \ce{ND} stretch & 2405 -- 2044 & 2352 -- 2061 \\ 
     \ce{COO} asymmetric stretch & 1604 & 1592 \\ 
     \ce{COO} symmetric stretch & 1408 & 1411 \\
     \ce{CH2} wagging & 1318 & 1326 \\
     \ce{ND3} asymmetric deformation & 1176 & 1175 \\
\hline
\end{tabular}
\vskip 0.5em
\parbox{\textwidth}{\centering\footnotesize
$^a$Experimental band positions are compared with literature values \citep{glyenviron,dglyref}.}
\end{table*}

The glycine-d$_3$ film was subsequently exposed to increasing fluences of 2\,keV electrons, up to a maximum fluence of $1.6\times10^{17}$\,\ce{e-}\,cm$^{-2}$. The resulting IR spectra are shown in Fig.~\ref{eirr}(b) and (c), with darker shades of green corresponding to higher electron fluences and the spectrum corresponding to the maximum fluence displayed with a black dashed line. Figure~\ref{eirr}(b) presents the spectra over the $2400-2100$\,cm$^{-1}$ region, which contains absorption features associated with \ce{CO2}, \ce{CO}, and \ce{OCN-}. These species are expected products of the energetic processing of glycine based on previous proton-irradiation experiments \citep{2026NatAs..10..531H} and can also be seen in the IR difference spectra shown in Appendix Fig.\ref{Diff_spec}. 
\ce{CO2}, a known destruction product of glycine \citep{protom,2014MNRAS.441.3209P}, has an absorption band at 2338\,cm$^{-1}$ that is observed after the first irradiation step and increases in intensity with increasing electron fluence. The \ce{CO2} is most likely formed through irradiation-induced decarboxylation of glycine \citep{Gerakines2012InSM}. As can be observed in both Figure~\ref{eirr}(b) and Appendix Fig.\ref{CO2_flux}, at higher irradiation fluences, the \ce{CO2} band starts to decrease in intensity, and two weaker bands appear at 2157 and 2136\,cm$^{-1}$, assigned to \ce{OCN-} and \ce{CO}, respectively. The \ce{CO} feature likely results, at least in part, from electron-induced decomposition of the \ce{CO2} formed during irradiation. \ce{CO} is a well-established product of energetic processing of \ce{CO2} \citep{Gerakines2012InSM,JHEETA2012208,UVCO2}, consistent with its emergence only at higher fluences after appreciable amounts of \ce{CO2} have accumulated. This would explain its detection at 2136\,cm$^{-1}$ and the observation of the abundance of \ce{CO2} decreasing at higher fluences. This trend is shown in Appendix Fig.\ref{CO2_flux}. The band at 2157\,cm$^{-1}$ is attributed primarily to the cyanate ion, \ce{OCN-} \citep{mate}, although contributions from other complex organic molecules containing related functional groups cannot be excluded. One possible contributor is \textit{N}-formylglycinamide (\ce{NH2CH2CONHCHO}), which was tentatively identified in previous experiments investigating proton irradiation of glycine \citep{2026NatAs..10..531H}. If electron irradiation drives similar reaction pathways, this species may also be formed under the present experimental conditions.

Figure~\ref{eirr}(c) shows the spectra over the wavenumber range $2700-2400$\,cm$^{-1}$. This region encompasses the broad absorption band associated with \ce{O-D} stretching modes \citep{OD}. An absorption therefore is expected here in these measurements following the formation of \ce{D2O}. As irradiation of the glycine-d$_{3}$ film progresses, the absorbance across this region increases. During initial irradiation, the absorbance increases gradually, followed by a more pronounced increase at intermediate electron fluences. At the highest fluences of greater than $2\times10^{16}$\,\ce{e-}\,cm$^{-2}$, the absorbance decreases with the final spectrum corresponding to $1.6\times10^{17}$\,\ce{e-}\,cm$^{-2}$ shown in a black dashed line. The \ce{D2O} formation and subsequent destruction can be seen plotted as a function of fluence in Appendix Fig.\ref{D2O_flux}.

The observed \ce{D2O} may be produced either through a direct, single-step condensation reaction or through a multi-step reaction pathway. The single-step condensation reaction, shown in Equation~\ref{pepformD3}, involves the formation of an amide bond between two glycine molecules, producing a peptide and a \ce{D2O} molecule. In contrast, multi-step reaction mechanisms would involve glycine destruction products generated during irradiation. These products could subsequently react to form \ce{D2O}, which would then be detected experimentally. One possible pathway could involve the reaction of an oxygen atom, produced during the irradiation-induced decomposition of \ce{CO2} to \ce{CO}, with a deuterium-bearing species. Such a multi-step mechanism would require a sufficient abundance of glycine destruction products and is therefore considered unlikely to account for the \ce{D2O} observed at low electron irradiation fluences, where the concentrations of these products are expected to be relatively low. The formation of \ce{D2O} at these low fluences, as shown in Appendix Fig.\ref{D2O_flux}, therefore supports its production through the single-step condensation reaction as shown in Equation \ref{pepformD3}. At higher electron irradiation fluences, the rate of \ce{D2O} formation decreases, and at the highest fluxes, the abundance of \ce{D2O} decreases. This could be due to sputtering \citep{Carmack_2024}, but also because, as the abundance of glycine destruction products increases, alternative multi-step reaction pathways may become more important and could produce other water isotopologues, such as \ce{HDO} and \ce{H2O} or lead to the destruction of the \ce{D2O}. Spectral confusion in the region containing the corresponding stretching modes, together with their lower production yields compared with those observed in previous proton irradiation experiments, made these species difficult to identify reliably in the IR spectra \citep{2026NatAs..10..531H}. Instead, QMS measurements during TPD following irradiation were used to determine whether \ce{HDO} and \ce{H2O} had also formed.

\begin{figure}[h]
     \resizebox{0.9\hsize}{!}{\includegraphics{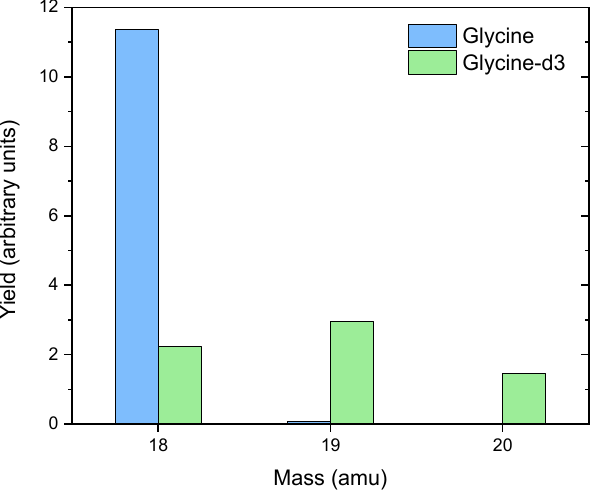}}
     \centering
     \caption{The yields at masses 18, 19, and 20\,amu, corresponding to \ce{H2O}, \ce{HDO}, and \ce{D2O}, respectively, measured during TPD following electron irradiation of glycine (blue) and partially deuterated glycine-d$_{3}$ (green).}
     \label{QMSelect}
 \end{figure}

Figure~\ref{QMSelect} shows the yields of masses 18, 19, and 20\,amu measured during a TPD experiment performed after the final electron irradiation step. The mass yields obtained for glycine are shown in blue, while those for glycine-d${3}$ are shown in green. The yield of each mass was calculated by integrating the QMS signal between 130 and 180\,K, corresponding to the temperature range over which water was expected to desorb from the substrate. The irradiated glycine film produced a strong signal at mass 18\,amu, with only negligible signals at 19 and 20\,amu, consistent with the formation of \ce{H2O}. In contrast, electron-irradiated glycine-d${3}$ produced significant signals at masses 18, 19, and 20\,amu, corresponding to \ce{H2O}, \ce{HDO}, and \ce{D2O}, respectively. The detection of \ce{D2O} is consistent with the increase in the \ce{O-D} stretching absorption observed during irradiation and shown in Fig.~\ref{eirr} and provides independent evidence that fully deuterated water is produced during electron processing.

The simultaneous detection of \ce{H2O} and \ce{HDO} indicates that the isotopic composition of the water is more complex than expected from the direct condensation reaction alone. However, the \ce{HDO} detected during TPD does not necessarily form during electron irradiation. Thermal H/D exchange between \ce{H2O}, \ce{D2O}, and perhaps glycine-d$_{3}$, can occur as the ice is heated, producing \ce{HDO} through proton/deuteron scrambling \citep{2015MNRAS.448.3820L}. \cite{2015MNRAS.448.3820L} demonstrated efficient H/D exchange in amorphous \ce{H2O}:\ce{D2O} ices over the temperature range $90-140$\,K, as well as H/D exchange involving other hydrogen-bonded species. Consequently, at least part of the mass 19\,amu signal observed here may be produced during the TPD itself rather than directly during electron irradiation. The \ce{D2O} detected by both IR spectroscopy during irradiation and QMS following irradiation provides the strongest evidence for the proposed condensation pathway. Formation of \ce{D2O} at low electron fluences, where the abundance of irradiation-induced destruction products is comparatively small, is consistent with a direct condensation reaction between glycine-d$_{3}$ molecules and the accompanying formation of a peptide bond. The similarity of this behaviour to that observed previously following proton irradiation of glycine, where ex-situ analysis unambiguously identified the dipeptide glycylglycine \citep{2026NatAs..10..531H}, supports the proposal that electron and proton processing can initiate comparable reaction pathways leading towards larger COMs and peptides.

\subsection{VUV measurements}

\begin{figure}[h]
     \resizebox{0.9\hsize}{!}{\includegraphics{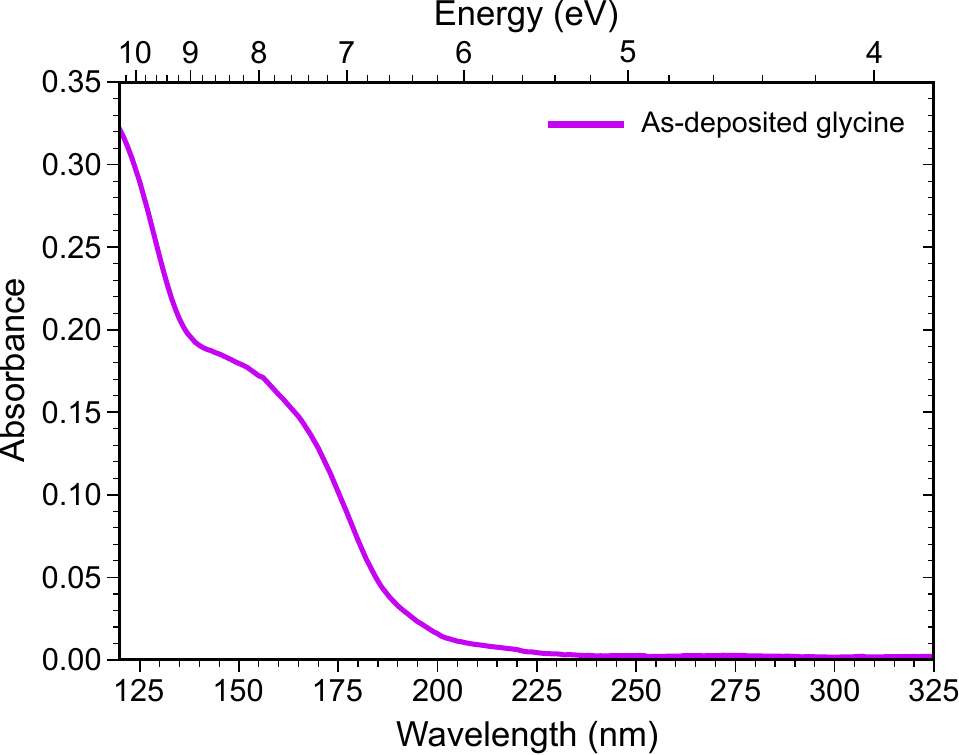}}
     \centering
     \caption{VUV spectrum of the initially deposited glycine with its characteristic band at 165\,nm and increasing absorbance towards and below 125\,nm.}
     \label{UVgly}
 \end{figure}

Figure~\ref{UVgly} shows the VUV spectrum of glycine deposited onto a \ce{MgF2} substrate at 20\,K. The spectrum exhibits two distinct features: an absorption band centred at approximately 165\,nm and a pronounced increase in absorbance towards wavelengths shorter than 125\,nm. The position and centre of the latter feature lie below the lower wavelength limit accessible with the experimental setup. Nevertheless, its presence is indicated by the continued increase in absorbance towards the short-wavelength detection limit. The as-deposited glycine VUV spectrum is broadly similar to the spectrum recorded in previous work \citep{UV_gly} but with slight differences. The spectrum here shows no absorption features at the higher wavelengths, which we attribute to amorphous glycine with no crystallinity, resulting from the low growth temperature of 20 K.

\begin{figure}[h]
     \resizebox{0.9\hsize}{!}{\includegraphics{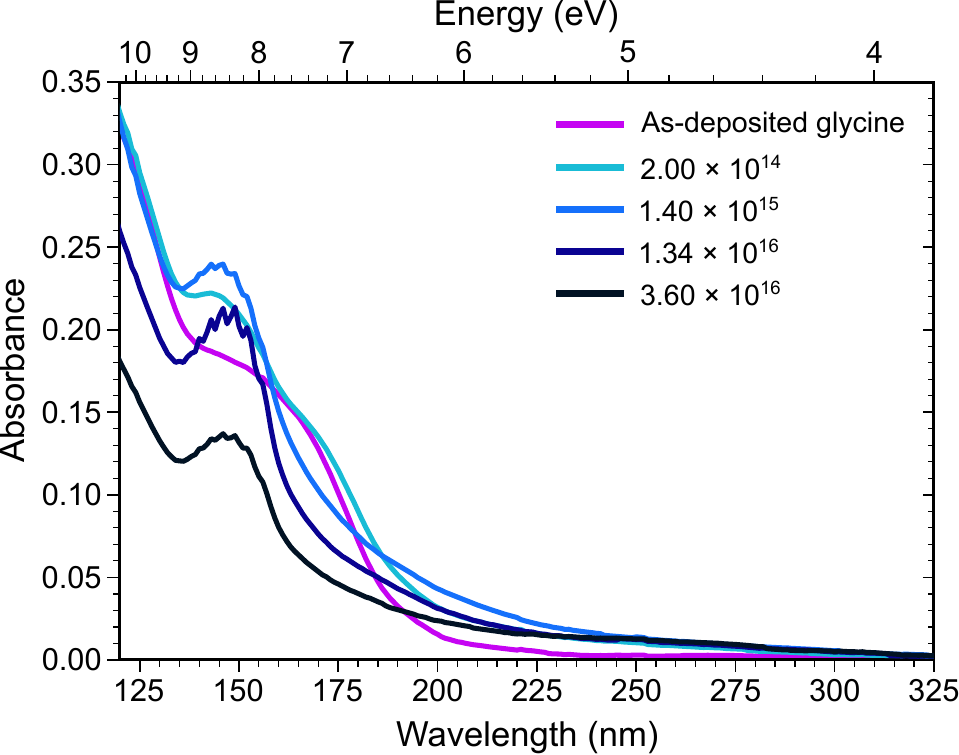}}
     \centering
     \caption{VUV spectrum of initially deposited glycine (lightest blue dashes), showing its characteristic absorption band at 165\,nm. Spectra recorded after successive electron irradiation steps are shown in progressively darker shades of blue, with the final step in black. The formation of \ce{CO} is indicated by the absorption feature near 150\,nm, with its characteristic substructure.}
     \label{UVirr}
 \end{figure}

Figure~\ref{UVirr} displays the VUV spectrum of unirradiated glycine (dashed line) together with spectra recorded after irradiation with increasing fluences of 1\,keV electrons (solid lines), up to a maximum fluence of $3.6\times10^{16}$\,\ce{e-}\,cm$^{-2}$. Spectra corresponding to progressively higher electron fluences are shown in increasingly dark shades of blue. As the electron fluence increases, the characteristic glycine absorption band at 165\,nm decreases in intensity and becomes difficult to distinguish after the first irradiation step. This behaviour indicates the progressive destruction of glycine molecules during electron bombardment. At the same time, new absorption features emerge during irradiation. In particular, the band centred at 147\,nm increases in absorbance with increasing electron fluence and develops a distinct substructure. Both the position and substructure of this feature are characteristic of \ce{CO} formation \citep{UVCO,UVwaterCOandCO2}. The formation of both \ce{CO2} and \ce{CO} during electron processing of glycine was independently demonstrated in Section~\ref{irsection}, where \ce{CO} formation was attributed to the electron-induced conversion of \ce{CO2} to \ce{CO} \citep{JHEETA2012208}. \ce{CO2} therefore is also likely to be present following irradiation; however, its contribution is more difficult to identify unambiguously in the VUV spectra because its absorption band occurs in a similar spectral region to that of \ce{CO}, but lacks the pronounced substructure characteristic of \ce{CO} \citep{UVwaterCOandCO2,UVCO2}. As the irradiation increases, there is increased absorption at the longer wavelengths greater than 200 nm. This could be due to the formation of larger COMs but also from changes in the optical properties of the ice. These include enhanced Rayleigh scattering that could be due to restructuring or sputtering of material by the electrons.

The VUV spectrum recorded after the final irradiation step, corresponding to a cumulative fluence of $3.6\times10^{16}$\,\ce{e-}\,cm$^{-2}$, is shown by the darkest blue trace in Fig.~\ref{UVirr}. At this fluence, the overall absorbance has decreased substantially. The glycine band at 165\,nm is no longer distinguishable, while the absorption towards 125\,nm is considerably weaker, consistent with extensive processing of the original glycine film. The feature at 147\,nm associated with \ce{CO} and \ce{CO2} also decreases in intensity at the highest fluence. This reduction may result from further electron-induced processing of these species, their removal from the surface through electron-stimulated desorption or sputtering, or the decreasing contribution from the underlying glycine absorption.

\begin{figure}[h]
     \resizebox{0.9\hsize}{!}{\includegraphics{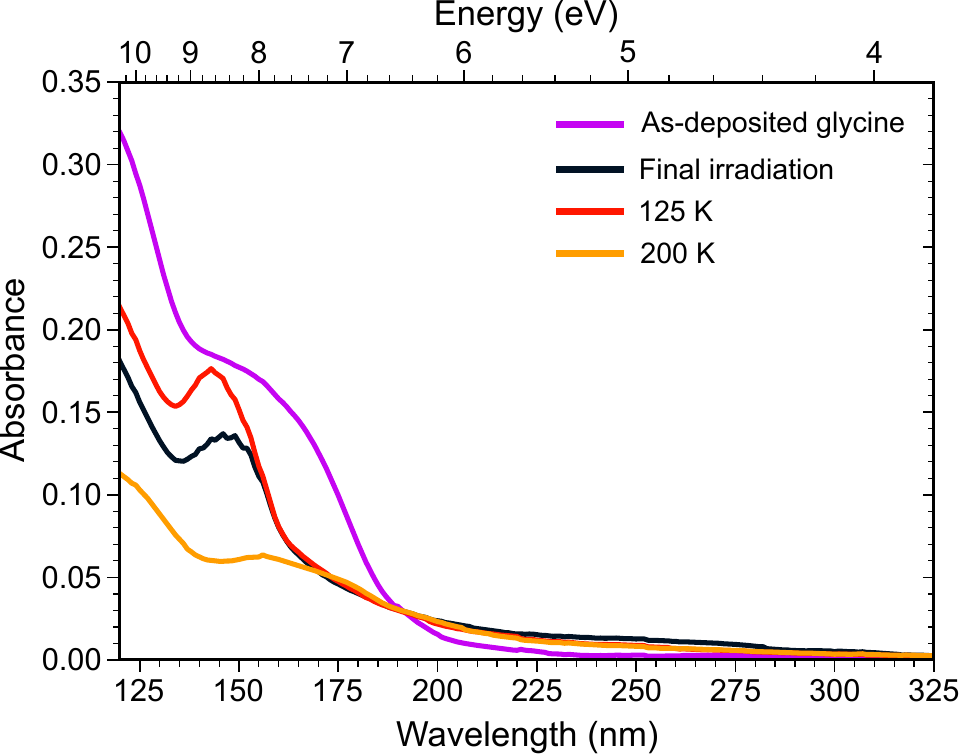}}
     \centering
     \caption{VUV spectrum of the as-deposited glycine (light blue dashes), fully irradiated glycine at 20\,K (black) compared to the same sample at 125\,K (red) and 200\,K (yellow).}
     \label{UVTPD}
 \end{figure}

Following the final irradiation step (black), the sample was heated to 125\,K to desorb volatile products, including \ce{CO2} and \ce{CO}. A VUV spectrum was subsequently recorded to characterise the less volatile material remaining on the surface and is shown in Fig.~\ref{UVTPD}. Figure~\ref{UVTPD} also includes the spectrum recorded after a further temperature ramp to 200\,K. The spectrum obtained immediately after the final electron irradiation step is shown in black, while the spectra recorded after heating to 125 and 200\,K are shown in red and yellow, respectively. Heating to 125\,K results in a substantial change in the VUV spectrum following the desorption of \ce{CO} and \ce{CO2}. In particular, a band centred at approximately 140\,nm becomes more prominent. Based on its position and spectral profile, this feature is assigned to \ce{H2O} \citep{UVwaterCOandCO2}. The increased prominence of the water band after heating may result from changes in the optical properties of the ice, including Rayleigh scattering associated with restructuring of the remaining water-containing material following the desorption of \ce{CO} and \ce{CO2} \citep{2021RSCAd..1133055J}. If the 140\,nm feature originates from water formed during electron irradiation of glycine, its intensity should decrease upon further heating as the water desorbs from the surface. Consistent with this interpretation, the spectrum recorded after heating to 200\,K no longer exhibits the 140\,nm band, and the absorbance in this spectral region is substantially reduced. The disappearance of this feature upon heating therefore supports its assignment to water and provides further evidence that \ce{H2O} is formed during the electron irradiation of glycine. 

The spectrum recorded after heating to 200\,K exhibits a broad absorption feature centred at approximately 160\,nm, together with an increasing absorbance towards wavelengths below 125\,nm. These features are similar to those observed in the spectrum of unprocessed glycine shown in Fig.~\ref{UVgly}, suggesting that a fraction of the deposited glycine remained unprocessed following electron irradiation. However, contributions from larger COMs formed during irradiation cannot be excluded, including peptide-like species produced through the energetic processing of glycine. This is observed more clearly in Appendix Fig.\ref{UV_residue}1. Overall, the VUV measurements provide an independent confirmation that water is formed during the electron irradiation of non-deuterated glycine, consistent with the behaviour observed in the IR and TPD experiments. 

Although water formation alone is not sufficient to demonstrate peptide formation, its detection strengthens the interpretation derived from the glycine-d$_{3}$ experiments, where \ce{D2O} is observed at low electron fluences and is consistent with the direct condensation reaction shown in Equation~\ref{pepformD3}. At higher fluences, multi-step reactions involving glycine destruction products are also likely to contribute to the water yield. 

Taken together, the IR, TPD, and VUV measurements therefore indicate that electron irradiation drives both glycine decomposition and condensation chemistry, with the latter providing a plausible pathway towards the formation of peptide bonds and larger COMs.

\section{Astrochemical implications}

The results presented here show that 1-2\,keV electron irradiation of glycine produces a chemical response similar to that previously observed following irradiation with 1\,MeV and 10\,keV protons \citep{2026NatAs..10..531H}. In both cases, glycine is destroyed to form species such as \ce{CO2}, \ce{CO}, and \ce{OCN-}, while water formation and a less volatile organic residue are also observed. These similarities support the hypothesis that low-energy secondary electrons generated during energetic processing play an important role in driving the subsequent chemistry, rather than the chemistry depending on the nature of the primary energetic source. Secondary electrons are expected to be produced in large numbers following the interaction of ionising radiation with interstellar ices and have been proposed as important drivers of radiation chemistry in these environments \citep{2019CSRev..48.2293A,2014FaDi..168..235M}.

The irradiation of partially deuterated glycine-d$_{3}$ provides evidence that condensation chemistry can occur during electron processing. The formation of \ce{D2O} at low electron fluences is consistent with a direct condensation reaction between glycine molecules, in which peptide-bond formation is accompanied by the release of \ce{D2O}. At higher fluences, the detection of \ce{H2O}, \ce{HDO}, and \ce{D2O} indicates that additional reaction pathways contribute to water formation. Although water production alone cannot provide unambiguous evidence for peptide formation, the combined IR, TPD, and VUV measurements are consistent with condensation chemistry occurring alongside glycine decomposition. These condensation reactions could be the formation mechanism that formed peptides in previous work \citep{2013ApJ...765..111K,Tushar}. In such experiments, simple interstellar ices are irradiated, which then form amino acids and peptides. It is therefore possible that, as in this paper and in \cite{2026NatAs..10..531H}, the formation mechanism for the peptides in these works was through condensation reactions.

The main astrochemical implication of these experimental results is that such chemistry may not require a specific energetic source. Cosmic rays penetrate dense molecular clouds and provide an important source of ionisation throughout regions where external UV radiation is strongly attenuated \citep{2013CSRev..42.7763I}. Their interaction with icy dust grains, together with other energetic radiation, can generate low-energy secondary electrons capable of initiating further chemistry within the ice \citep{2016ChRv..116.9631O,2019CSRev..48.2293A,2014FaDi..168..235M}. The present results therefore suggest that energetic processing of amino acids should not be viewed solely as a destruction mechanism, but may also provide a pathway towards greater molecular complexity, including the possible formation of peptide bonds, across a range of interstellar and circumstellar environments.

This chemistry may have consequences for the molecular inventory inherited by planetary systems. Observations of comet 67P indicate that at least part of its volatile material predates Solar System formation and may preserve an interstellar chemical heritage \citep{2020SSRv..216..102R}, while the recent detection of glycine and numerous other amino acids in returned samples from asteroid Bennu demonstrates that amino-acid-rich material can be incorporated into primitive Solar System bodies \citep{2025NatAs...9..199G}. Energetic processing of amino acids before or during their incorporation into such bodies could therefore increase the complexity of the organic material ultimately delivered to planetary surfaces. The present experiments support low-energy secondary-electron-driven chemistry as a potentially widespread pathway linking interstellar amino acids to more complex prebiotic molecules.

\begin{acknowledgements}
Support from the Danish National Research Foundation through the Centre of Excellence `InterCat' (grant agreement No. DNRF150), and the Europlanet 2024 RI, which was funded by the European Union Horizon 2020 Research Innovation Programme under grant agreement No. 871149 is acknowledged. The main components of the ICA setup were purchased using funds obtained from the Royal Society through grants UF130409, RGF/EA/180306, and URF/R/191018. Further developments of the installation were supported in part by the Eötvös Loránd Research Network through grants ELKH IF-2/2019 and ELKH IF-5/2020. Support has also been received from the Research, Development, and Innovation Fund of Hungary through grant nos. K128621 and ADVANCED-151196. This paper is also based on work from the COST Actions CA20129 “MultIChem” and CA22133 “PLANETS”, supported by COST (European Cooperation in Science and Technology).

\end{acknowledgements}
\bibliographystyle{aa} 
\bibliography{Bib}
%
%

\begin{appendix}

\section{}\label{IR_Diff}

\begin{figure}[h]
     \resizebox{0.9\hsize}{!}{\includegraphics{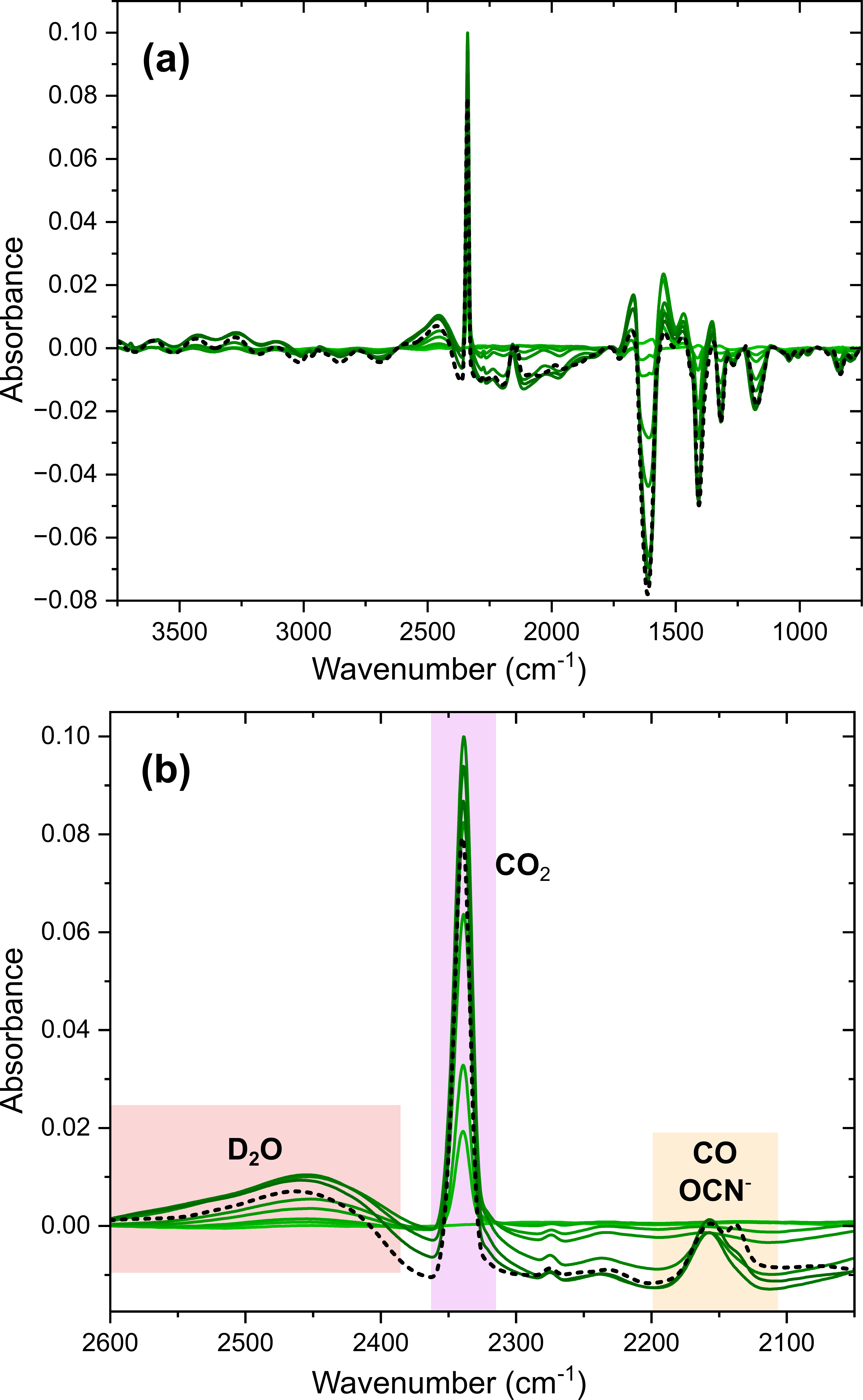}}
     \centering
     \caption{IR difference spectra where the recorded spectrum has the as-deposited spectrum subtracted from it for each electron irradiation step, where the darker the colour, the larger the fluence of electrons and the dashed black spectra are the final measurement. (a) shows the whole spectral range, with negative features showing where bands have decreased and correspond to partially deuterated glycine-d$_{3}$. Positive features are bands that are detected as a result of the formation of new molecules. (b) shows the region from 2600 to 2050\,cm$^{-1}$ focusing on new features corresponding to the formed \ce{D2O} (red highlight), \ce{CO2} (purple highlight), \ce{CO}, and \ce{OCN-} (orange highlight).}
     \label{Diff_spec}
 \end{figure}

Figure \ref{eirr} in the main text shows the IR spectra for partially deuterated glycine-d$_{3}$ and focuses on particular wavelength regions after being irradiated with electrons. Figure \ref{Diff_spec} shows the same spectra, but each has had the as-deposited spectrum of glycine-d$_{3}$ subtracted from it to show changes caused by the electrons. Figure \ref{Diff_spec}(a) shows the full wavelength range where negative absorption bands demonstrate that glycine-d$_{3}$ has been processed, so it is less present on the surface. Figure \ref{Diff_spec}(b) focuses on the wavelength region from 2600 to 2050\,cm$^{-1}$, where the same bands as were seen previously, those corresponding to \ce{D2O}, \ce{CO2}, \ce{CO}, and \ce{OCN-}, can be seen to have increased after irradiation.

\section{}\label{IR_fluence}

\begin{figure}[h]
     \resizebox{0.9\hsize}{!}{\includegraphics{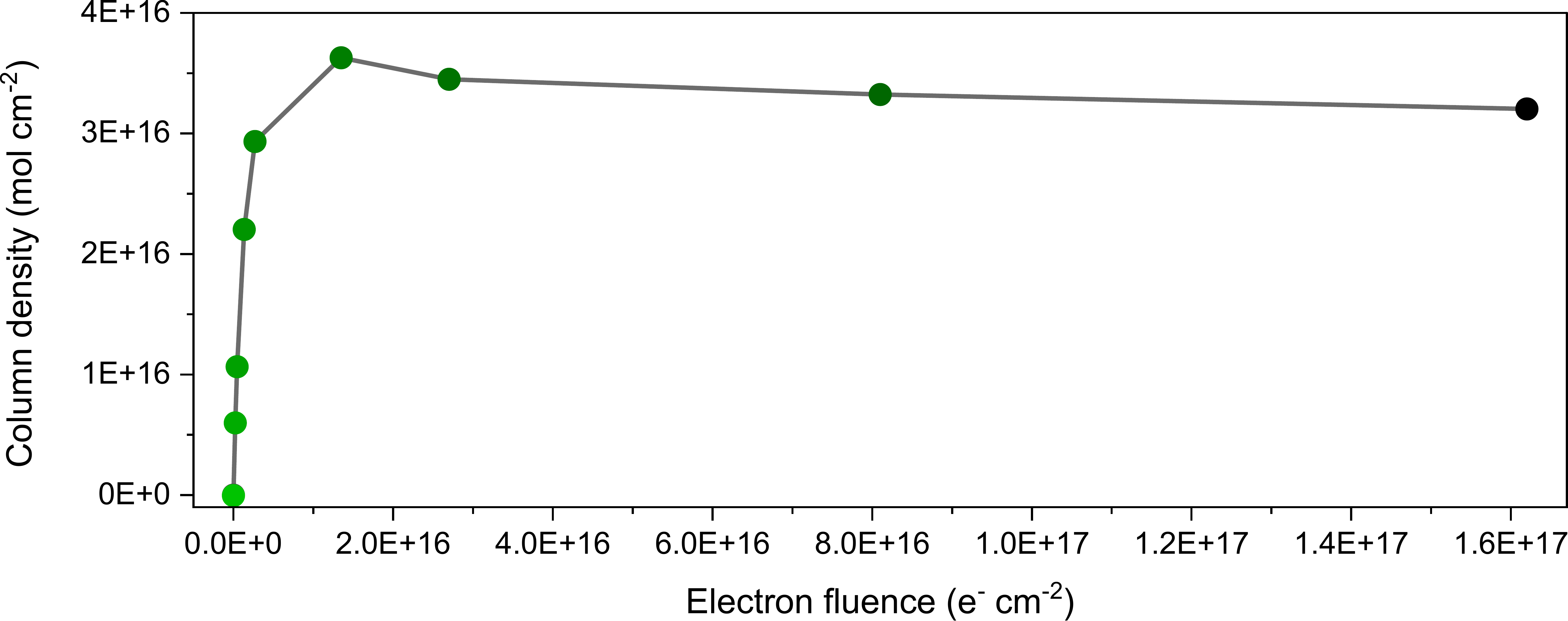}}
     \centering
     \caption{Column density of \ce{CO2} as a function of electron irradiation fluence. \ce{CO2} is formed quickly with low electron fluences as the deposited glycine is destroyed as it undergoes decarboxylation. At electron fluences greater than $2.0\times10^{16}$\,\ce{e-}\,cm$^{-2}$ the column density decreases. This is a combination of the effects of the processing of \ce{CO2} into \ce{CO} and sputtering.}
     \label{CO2_flux}
 \end{figure}

Figure \ref{CO2_flux} shows the column density of \ce{CO2} as a function of electron irradiation. The column density was calculated using the area of the \ce{CO2} asymmetric stretching mode at 2338\,cm$^{-1}$ and the band strength of $8.78\times10^{-17}$ cm per molecule, which was calculated in \citep{hudson2025carbon}. It shows that \ce{CO2} is formed quickly even at low electron irradiation fluences. This is due to the destruction of glycine and its decarboxylation, which is known to occur when glycine is energetically processed. At larger electron fluences of greater than $2.0\times10^{16}$\,\ce{e-}\,cm$^{-2}$ the column density of \ce{CO2} no longer increases and instead starts to decrease. This is likely due to a combination of sputtering and the electron-induced decomposition of \ce{CO2} to form \ce{CO}.

\begin{figure}[h]
     \resizebox{0.9\hsize}{!}{\includegraphics{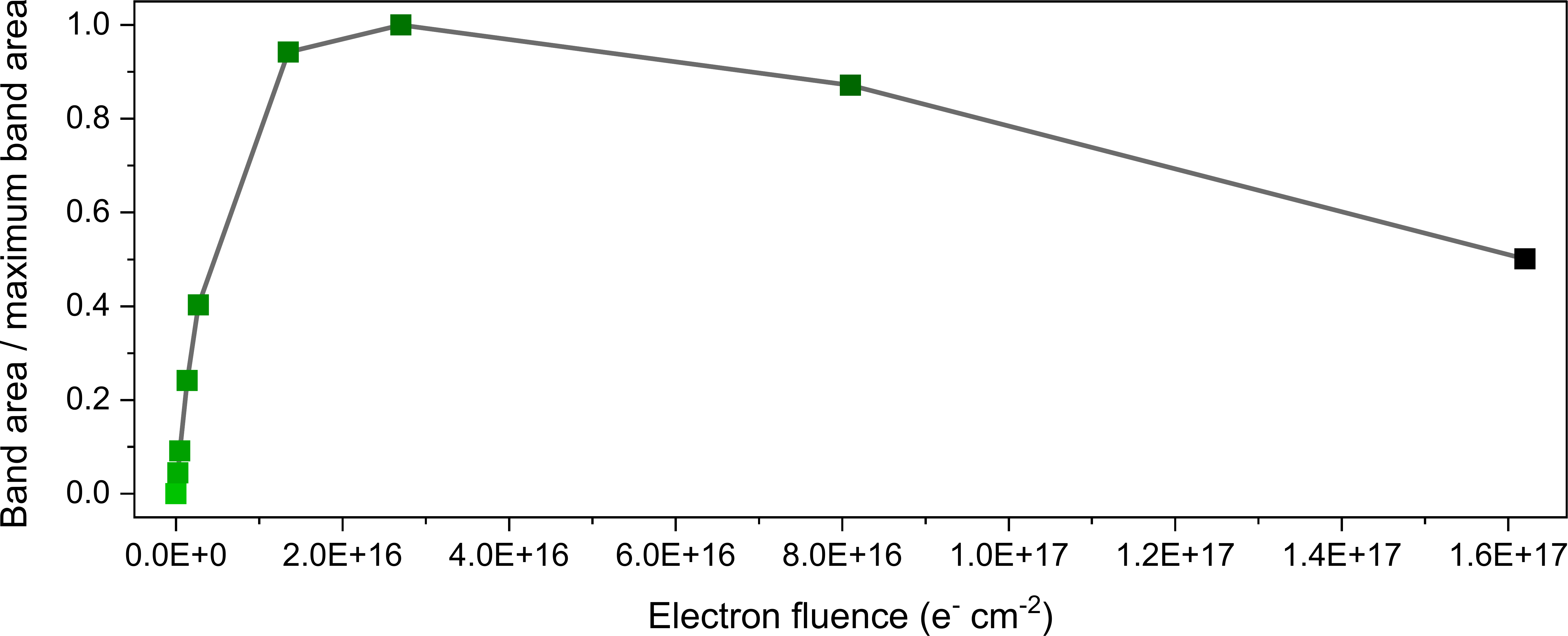}}
     \centering
     \caption{The band area between 2630 and 2400\,cm$^{-1}$, which corresponds to the \ce{O-D} stretching modes of \ce{D2O}, normalised to the maximum area of the band, plotted as a function of electron fluence. This shows that \ce{D2O} is formed rapidly at low electron fluences, likely due to single-step condensation reactions before its formation rate slows and decreases at large irradiation steps.}
     \label{D2O_flux}
 \end{figure}

Figure \ref{D2O_flux} shows the integrated band area between 2630 and 2400\,cm$^{-1}$ normalised to the largest measured area. This region contains the \ce{O-D} stretching modes of \ce{D2O}, but the absorption feature stretches to lower wavenumbers also. However, spectral confusion and overlap between absorption features for glycine-d$_{3}$ and \ce{D2O} mean that the whole band cannot be measured and therefore the figure is not plotted as a function of column density. At the lower electron fluences, the area of the band and therefore the amount of formed \ce{D2O} increases. This is consistent with previous measurements \citep{2026NatAs..10..531H} where single-step condensation reactions to form amide bands produce \ce{D2O} as a secondary product. At larger electron fluences, the area of this band starts to decrease. This was not observed previously and is attributed to a combination of further electron-induced reactions and sputtering.

\section{}\label{UV_residue}

Figure \ref{UVTPD} shows that, at 200 K, material remains on the surface. This spectrum has similarities with that of the as-deposited glycine and likely contains some unreacted glycine. However, there are noticeable differences between the two spectra, which could indicate that different, larger COMs are also present. To demonstrate this, the as-deposited spectrum was scaled by 0.35 such that it and the 200 K spectrum have the same absorbance at 125 nm. The scaled as-deposited spectrum was then subtracted from the 200 K spectrum to create a difference spectrum, which can be seen in Figure \ref{UV_residue}1.

\begin{figure}[h]
     \resizebox{0.9\hsize}{!}{\includegraphics{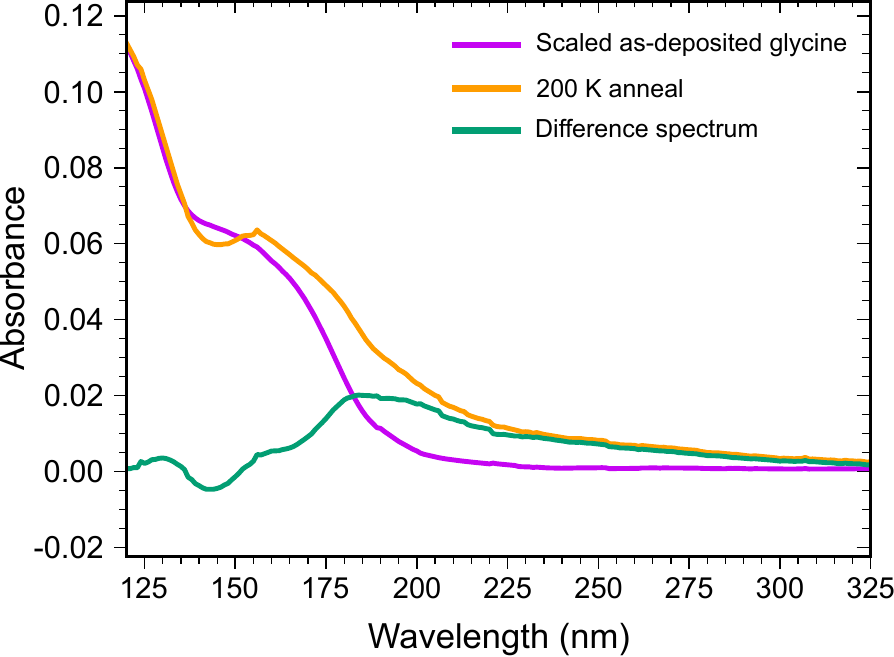}}
     \centering
     \caption{VUV spectrum of the as-deposited glycine scaled by 0.35 (light blue dashes), fully irradiated glycine heated to 200\,K (yellow), and a difference spectrum (green) showing the changes between the two previous spectra.}
     \label{Residual_UV}
 \end{figure}

This difference spectrum highlights some differences between the two spectra. Of note is the increased absorbance at wavelengths above 175 nm. As-deposited glycine has no absorption here and therefore could be due to the formation of other COMs. This absorption alone is not enough to identify larger possible COMs and could be due to other effects such as Rayleigh scattering due to ice restructuring or electron sputtering

\end{appendix}

\end{document}